\documentclass[11pt]{article}
\usepackage{amssymb, latexsym, amsmath, amsthm, amsfonts, verbatim, bm}
\usepackage[final]{graphicx}
\usepackage{xcolor}
\usepackage[pdf]{pstricks} 
\usepackage{pst-plot}
\usepackage{hyperref}
\begin{document}
\newcommand{\B}[1]{\textbf{#1}}
\newcommand{\ra}{\rightarrow}
\newcommand{\sinc}{\text{sinc}}
\newcommand{\supp}{\text{supp}}
\newcommand{\suppe}{\text{supp}_{\varepsilon}}
\newcommand{\ind}{\B{1}}
\newcommand{\sgn}{\text{sgn}}
\newcommand{\etal}{\text{et al. }}
\newcommand{\Beta}{{\boldsymbol\beta}}
\newcommand{\Radon}{\mathcal{R}}
\newcommand{\Xray}{\mathcal{P}}
\newcommand{\Fourier}{\mathcal{F}}
\newcommand{\Hilbert}{\mathcal{H}}
\newcommand{\Identity}{\mathcal{I}}
\newcommand{\BL}{\mathcal{L}}
\newcommand{\R}{\mathcal{R}}
\newcommand{\Cov}{\text{Cov}}
\newcommand{\argmin}{\operatorname{argmin}}
\title{Tomographic Model Based Iterative Reconstruction\\ of Symmetric Objects}
\author{Kyle M. Champley, Ibrahim Oksuz, Matthew G. Bisbee, \\ Joseph W. Tringe, and Brian Maddox}
\maketitle

\begin{abstract}
\noindent Computed Tomography (CT) reconstruction of objects with cylindrical symmetry can be performed with a single projection.  When the measured rays are parallel, and the axis of symmetry is perpendicular to the optical axis, the data can be modeled with the so-called Abel Transform.  The Abel Transform has been extensively studied and many methods exist for accurate reconstruction.  However, most CT geometries are cone-beam rather than parallel-beam.  Using Abel methods for reconstruction in these cases can lead to distortions and reconstruction artifacts.  Here, we develop analytic and model-based iterative reconstruction (MBIR) methods to reconstruct symmetric objects with an arbitrary axis of symmetry from a cone-beam geometry.  The MBIR methods demonstrate superior results relative to the analytic inversion methods by mitigating artifacts and reducing noise while retaining fine image features.  We demonstrate the efficacy of our methods using simulated and experimentally-acquired x-ray and neutron projections.
\end{abstract}

\section{Introduction}

Computed Tomography (CT) is a nondestructive and noninvasive three-dimensional imaging modality used in medicine, industrial manufacturing, and security screening \cite{MartzBook}.  Since its inception in the 1970s \cite{RiviereCrawford2021, Boone2021}, CT has revolutionized medicine by eliminating exploratory surgery and providing clinicians the ability to quickly diagnose disease.  Industrial manufacturing employs CT to inspect critical parts for flaws and defects.  CT is also employed at airports and seaports to inspect cargo and luggage for explosives and other contraband.  New applications for CT continue to drive development of advanced imaging systems and data processing/ reconstruction algorithms.
Various forms of penetrating radiation including x-rays or neutrons can be used for computed tomography.  Typical industrial CT scanners employ a source of the penetrating radiation opposite an area detector; the object is placed on a rotation stage between the source and detector.  Projections are acquired over 180$^\circ$ or more, and this data is used in tomographic reconstruction algorithms \cite{Nat86} to form a 3D map of the Linear Attenuation Coefficient (LAC) distribution of the object.

This paper is focused on reconstruction of objects that are cylindrically symmetric using only a single projection.  Applications for this approach include explosive dynamics and shock physics experiments \cite{Scarpetti_FXR_1997, Multhauf_FXR_2005, Smith_PPPC_2005, Scarpetti_PAC_2007, Nielsen2018}, fluid dynamics \cite{Sznitman2007}, and pyrometry \cite{Dreyer2019}.  In many cases the object being imaged is evolving over short times where it is only possible to collect a single projection.  After reconstruction, details which are impossible to differentiate in a single radiograph can often be seen clearly in 3D.

When the radiation source is far from the object being imaged, then the beam can be approximated by a 2D set of parallel rays.  In addition, if the object being imaged is cylindrically symmetric with the axis of symmetry perpendicular to the optical axis, then data can be modeled by the so-called Abel Transform.  Analytic inversion of the Abel Transform has been extensively studied and is the most common approach for reconstructing symmetric objects from a single projection \cite{PyAbel}.  The Abel inversion has several shortcomings, however.  It assumes a limited beam geometry of parallel rays, and that the axis of symmetry be perpendicular to the optical axis; and it does not include a model of the noise statistics.  A few authors have studied model-based iterative reconstruction methods \cite{Asaki_IP_2005, Abramson_SIAM_2008, Howard_SIAM_2016} to reconstruct with the Abel transform, but it is unclear how the optimization is performed because there has been no explicit mention of the adjoint operator.

Here we will develop reconstruction capabilities that overcome the noted drawbacks of the inversion of the Abel transform.  The work is motivated by the development of Model Based Iterative Reconstruction (MBIR) algorithms developed for conventional CT \cite{Fessler_PCG_1999,Bouman_MedPhys_2007,Fu_PCG_2012}.

There are several software packages available for reconstruction of symmetric objects from a single projection.  The PyAbel \cite{PyAbel} open-source Python software package contains several different implementations of the forward and inverse Abel transforms.  The Bayesian Inference Engine (BIE) \cite{Cunningham_BIE_1996}, which is written in Smalltalk and C/C++/Fortran, features a flexible user interface that enables users to assemble physics-based optimization algorithms in a GUI-focused toolkit.  The Livermore Inference Engine (LIE) \cite{Maddox_LIE_2021} is a flexible Python toolkit that uses a Bayesian framework for model-based reconstruction of radiographic data similar to BIE.  LIE enables one to quickly implement new models and reconstruction methods, leverage advances in machine learning, and can take better advantage of modern supercomputing resources.

The algorithms developed here are available in the Livermore Tomography Tools (LTT) software package \cite{LTTpaper} and the LEAP-CT \cite{LEAPpaper} open source software package.  These libraries provide Python interfaces that allows new tools to be developed that can access many of the LTT and LEAP-CT algorithms.  Users of LIE can leverage LTT or LEAP-CT projectors, backprojectors, and other algorithms through this Python interface.  LTT and LIE are available upon request and LEAP-CT is available on github at: \href{https://github.com/LLNL/LEAP/}{https://github.com/LLNL/LEAP/}.

\section{X-ray and Abel Transforms and Assumptions of Symmetry}

The X-ray Transform of a function $f \in L^1(\mathbb{R}^3)$ is given by
\begin{eqnarray}
\mathcal{P}f(\B{y}, \Theta) &:=& \int_\mathbb{R} f(\B{y} + t\Theta) \, dt,
\end{eqnarray}
where $\B{y}, \Theta \in \mathbb{R}^3$.  In a typical data collection configuration, a point source illuminates an area detector. This leads to the so-called cone-beam coordinates which is given by
\begin{eqnarray}
\mathcal{P}f(\varphi, u, v) &:=& \int_\mathbb{R} f\left(R\bm{\theta}(\varphi) - \tau\bm{\theta}^\perp(\varphi) + \frac{l}{\sqrt{D^2 + u^2 + v^2}} \left[ -D\bm{\theta}(\varphi) + u\bm{\theta}^\perp(\varphi) + v\widehat{\B{z}} \right] \right) \, dl,
\end{eqnarray}
where $\bm{\theta}(\varphi) = (\cos\varphi, \sin\varphi, 0)^T$, $\bm{\theta}^\perp(\varphi) = (-\sin\varphi, \cos\varphi, 0)^T$, and $\widehat{\B{z}} = (0,0,1)^T$.  Here, $R$ is the distance from the source to the center of rotation, $\tau$ is a horizontal offset of the source (also can be used to specify a detector rotation around detector column axis), $\varphi$ is the rotation angle of the object, $D$ is the source to detector distance, and $(u, v)$ are the detector column and row coordinates, respectively.

Let the Abel Transform of a cylindrically-symmetric function $f \in L^1[0,\infty)$ be denoted by
\begin{eqnarray}
\mathcal{A}f(y) := 2\int_y^\infty f(r) \frac{r}{\sqrt{r^2-y^2}} \, dr
\end{eqnarray}
In the following we shall denote the discrete versions of $\mathcal{P}$ and $\mathcal{A}$ by $P$ and $A$, respectively.

We assume that the object being imaged is cylindrically symmetric and has axis of symmetry given by $(\sin\alpha, 0, \cos\alpha)^T$ and a single projection at $\mathcal{P}f(0, u, v)$.  For objects whose axis of symmetry is not perpendicular to the y-axis, we assume that the projection image can be rotated such that the axis of symmetry can be described as previously stated.  The axis of symmetry may not be aligned with the z-axis for several reasons.  First, a tilted axis of symmetry may be an unintentional misalignment of the experiment.  Second, one may deliberately tilt the axis of symmetry to avoid a penetration of a dense line of sight, for similar reason as computed laminography; this may prevent accurate reconstruction.  One may also deliberately tilt the axis of symmetry to align the system with an important planar feature of interest.  This may be necessary because, to recover a feature in CT, one must collect a projection that is essentially tangent to the feature.  In most cases it is desirable that $|\alpha| \leq 10^\circ$ to avoid significant cone-beam artifacts \cite{Leach_NDTE_2021}.

\section{Model-Based Iterative Reconstruction}

Model-Based Iterative Reconstruction in x-ray CT entails solving the following numerical optimization problem
\begin{eqnarray}
\widehat{f} &:=& \argmin_f \Phi(f|g) + \beta \Psi(f),
\end{eqnarray}
where $\Phi$ is a data fidelity term of the difference between the measurements, $g$, and the forward projections of the volume, $f$, and $\Psi$ is a prior term (regularizer) on the space of solutions, $f$, which encourages smoothness of the solution.  The strength of the regularization functional is controlled by $\beta \geq 0$.  The $\Psi$ term is often of Total Variation (TV) type \cite{ROF_TV_1992,Yu_LineSearch_2006}.  A common data fidelity term is a squared Weighted Least Squares (WLS) norm which is given by
\begin{eqnarray}
\Phi(f|g) &:=& \frac{1}{2}\left(Pf - g\right)^T W \left(Pf - g\right).
\end{eqnarray}
This loss term models the noise of $g$ as Gaussian with covariance matrix given by $W^{-1}$.  Solving such cost functions requires one to calculate the forward as well as adjoint operators.  Note that for real discrete operators the adjoint is equal to the matrix transpose.  For example, when one uses the WLS data fidelity term, the gradient is equal to
\begin{eqnarray}
\Phi'(f|g) = P^*W(Pf - g).
\end{eqnarray}
The adjoint of the X-ray Transform for symmetric objects and the Abel Transform are rarely discussed in the literature, yet are essential for reconstruction.

The cost function with WLS data fidelity and regularization term is referred to as Regularized Weighted Least Squares (RWLS).  This is the cost function that we demonstrate for our numerical experiments in this work, but once one has the forward and backprojection methods, it should possible to easily implement any MBIR algorithm.

\subsection{Adjoint Operators}

Model based iterative reconstruction algorithms of linear operators such as the X-ray and Abel Transforms require the computation of the forward operator and its adjoint.  The adjoint of a linear operator, $L$, is defined as the transform, $L^*$, that solves the identity
\begin{eqnarray}
<Lf, g> = <f, L^* g>. \label{adjointDefinition}
\end{eqnarray}
Thus one can show that the adjoint of the cone-beam X-ray Transform and the Abel Transform are given by
\begin{eqnarray}
\mathcal{P}^*g(\B{x}) &=& \frac{1}{D^2}\int \frac{\sqrt{D^2 + u^2(\B{x}, \varphi) + v^2(\B{x}, \varphi)}}{\left( R - \B{x} \cdot \bm{\theta}(\varphi) \right)^2} g(\varphi, u(\B{x}, \varphi), v(\B{x}, \varphi)) \, d\varphi \\
u(\B{x}, \varphi) &:=& D\frac{\B{x} \cdot \bm{\theta}^\perp(\varphi) + \tau}{R - \B{x} \cdot \bm{\theta}(\varphi)} \label{eq:backprojection_u} \\
v(\B{x}, \varphi) &:=& D\frac{x_3}{R - \B{x} \cdot \bm{\theta}(\varphi)} \label{eq:backprojection_v}
\end{eqnarray}
and
\begin{eqnarray*}
\mathcal{A}^* g(r) &=& 2r \int_0^r \frac{g(y)}{\sqrt{r^2-y^2}} \, dy,
\end{eqnarray*}
respectively.  Both of these adjoints may be found using equation (\ref{adjointDefinition}) and a change of variables.  Note that the equations for $u(\B{x}, \varphi)$ and $v(\B{x}, \varphi)$ are found by solving
\begin{eqnarray*}
\B{x} = R\bm{\theta}(\varphi) - \tau\bm{\theta}^\perp(\varphi) + \frac{l}{\sqrt{D^2 + u^2 + v^2}} \left[ -D\bm{\theta}(\varphi) + u\bm{\theta}^\perp(\varphi) + v\widehat{\B{z}} \right]
\end{eqnarray*}
for $u$ and $v$.

\begin{figure}[h!]
\begin{center}
\begin{tabular}{cc}
\includegraphics[width=0.45\textwidth]{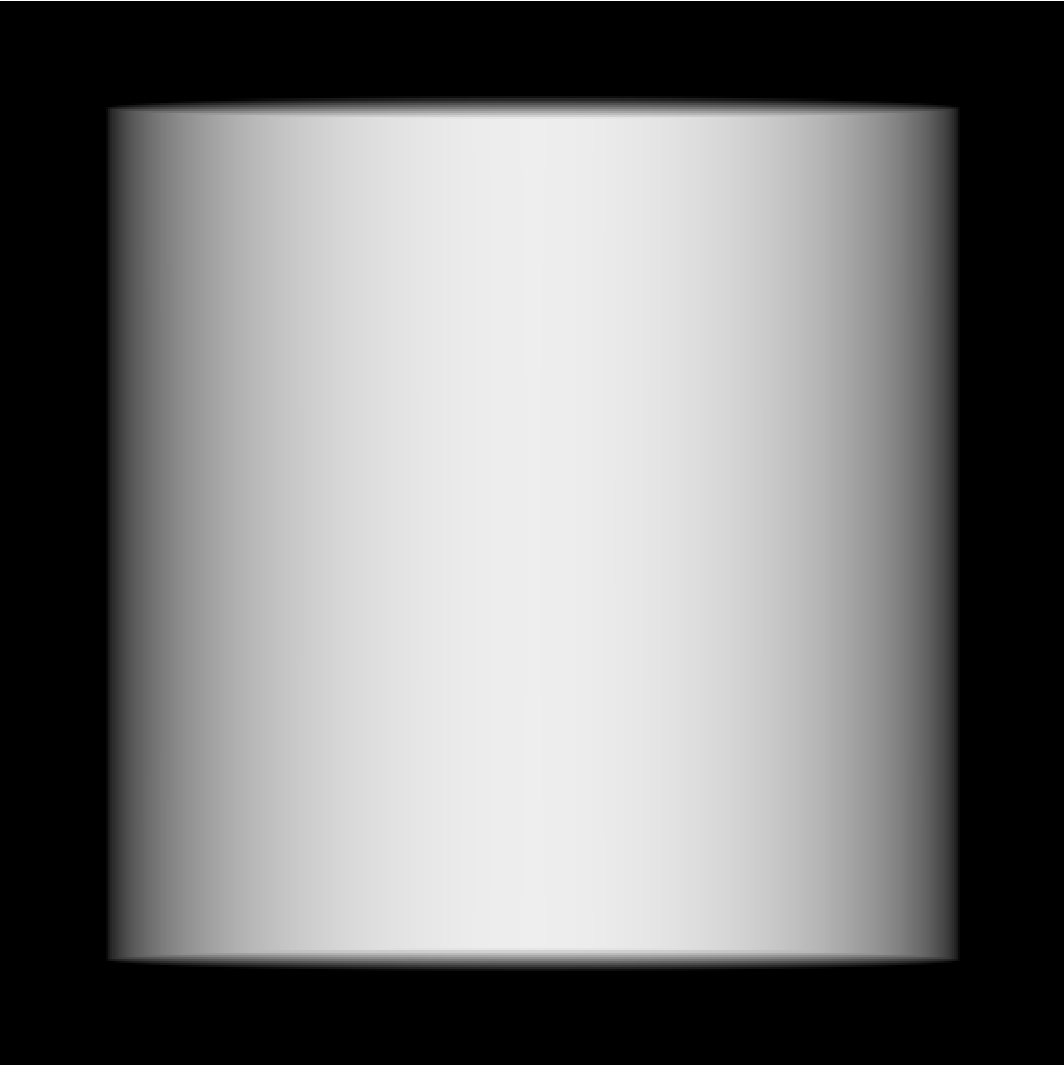}
& \includegraphics[width=0.45\textwidth]{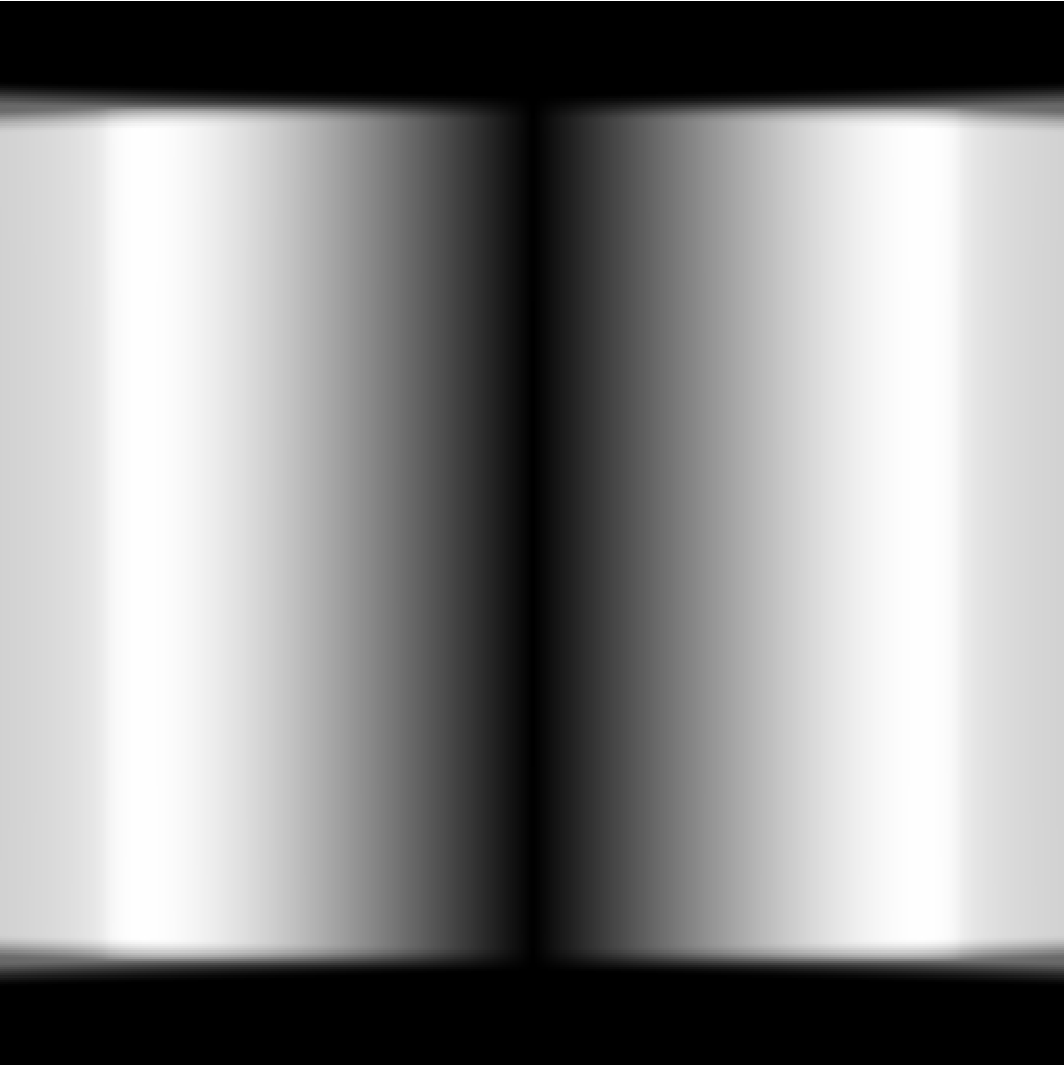}
\end{tabular}
\end{center}
\caption{Backprojection of a single projection of the (left) standard and (right) symmetric cone-beam X-ray Transform.  The object here is a uniform cylinder and the images show a vertical two-dimensional slice through the center of the object.} \label{fig:adjoint}
\end{figure}

Although both the X-ray and Abel Transforms are line integral transforms, their adjoints have distinctly different weighting terms inside the integrals.  The most notable difference is that the adjoint of the Abel transform has a singularity in its integrand.  See Figure \ref{fig:adjoint} for backprojection images of the standard and symmetric cone-beam X-ray Transforms.  One can clearly see the singularity along the axis of symmetry.  The pixels closest to the axis of symmetry are close to, but not equal to, zero.  We also note that the adjoint of the X-ray Transform is often referred to as the backprojection operator.  In this paper we will use the terms adjoint and backprojection interchangeably.

In practice, it is not necessary to analytically derive the adjoints of these transforms.  One can simply use the same code that calculates the entires of the system matrix in the forward projection for the adjoint (backprojection).  Regardless, it is useful to understand the features of the adjoints when implementing MBIR algorithms.

\begin{figure}[h!]
\begin{center}
\begin{tabular}{cc}
\includegraphics[width=0.45\textwidth]{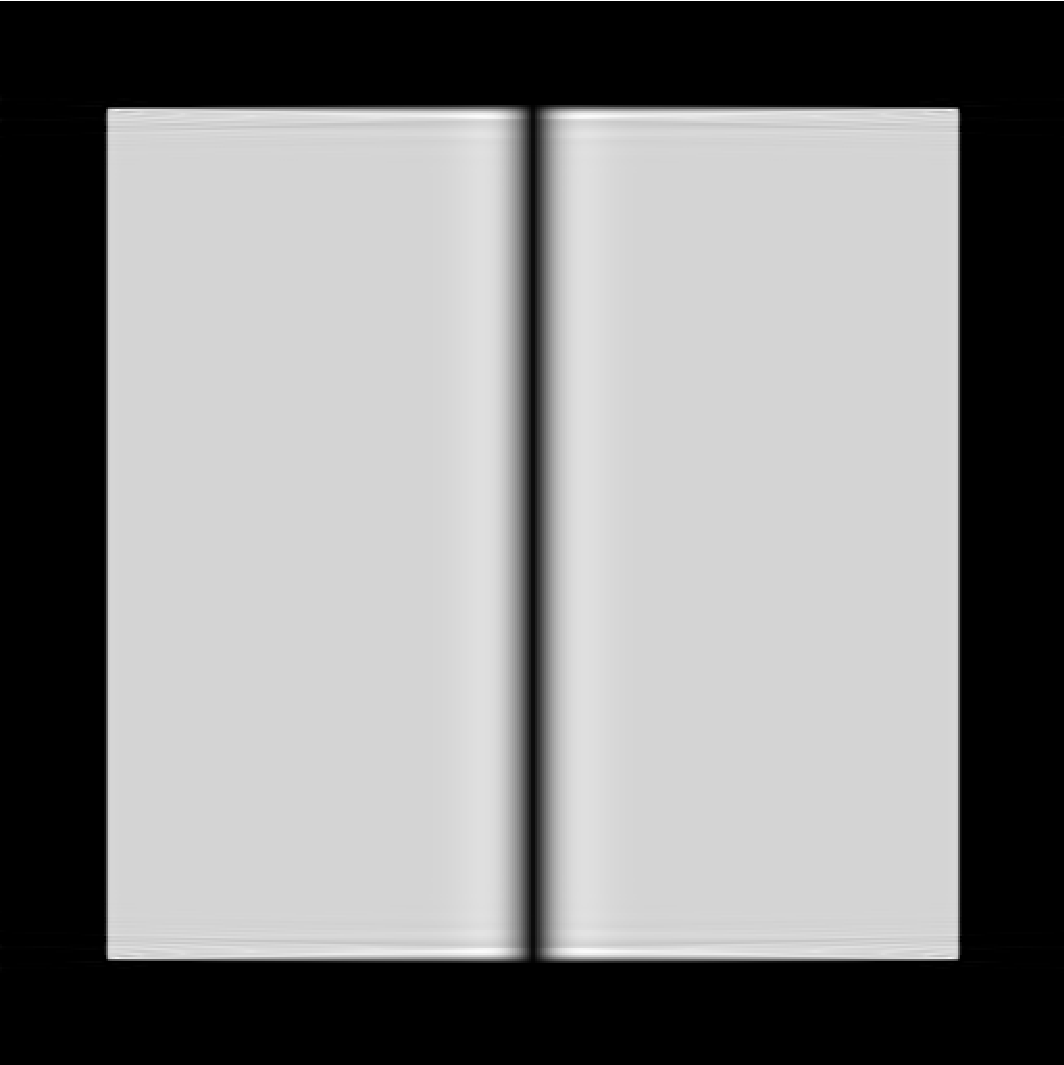}
& \includegraphics[width=0.45\textwidth]{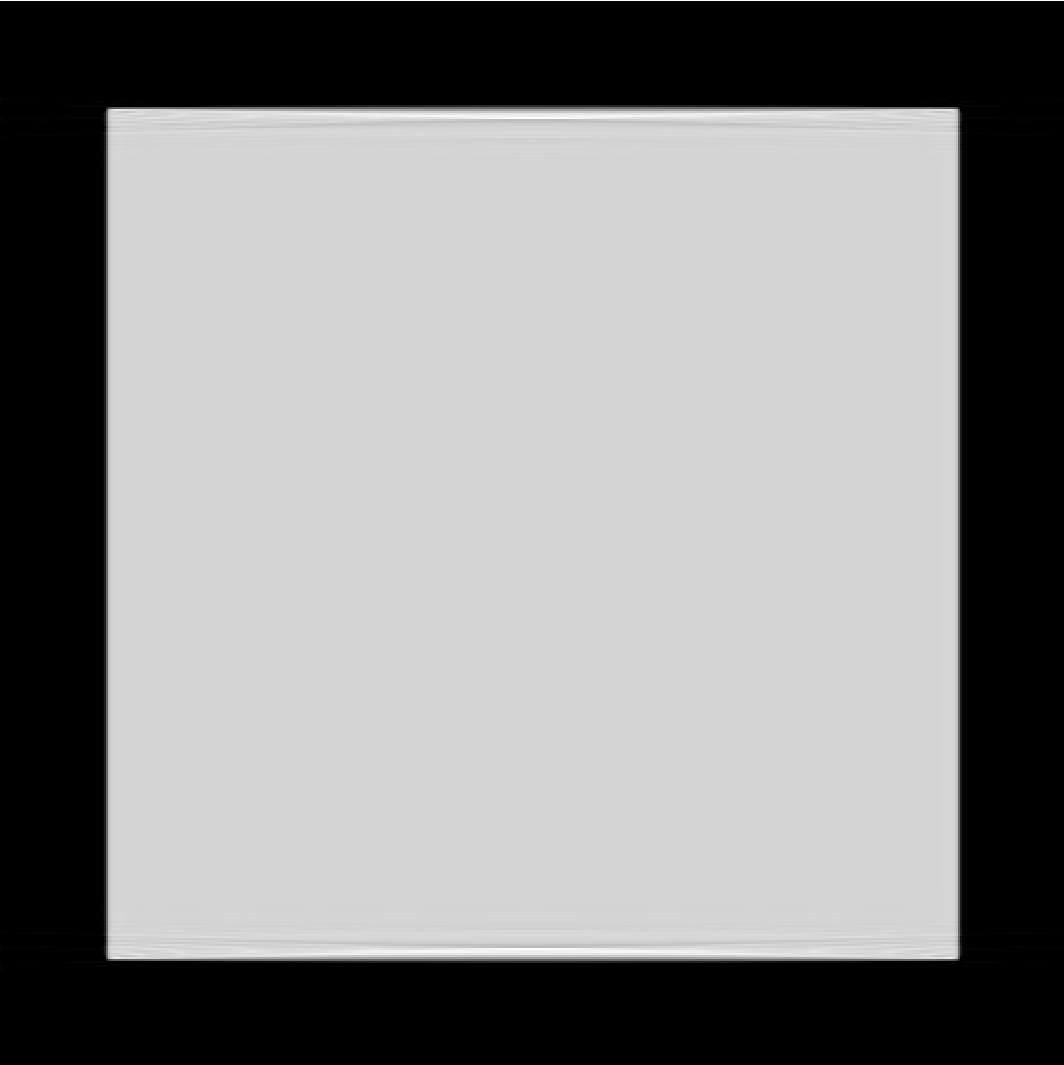}
\end{tabular}
\end{center}
\caption{Vertical slices of reconstructions of a uniform cylinder using the symmetric cone-beam transform using one hundred iterations of conjugate gradient.  The image on the right uses the SQS preconditioner while the image on the left does not.} \label{fig:preconditionerEffect}
\end{figure}

When implementing MBIR algorithms for symmetric objects, pixels near the axis of symmetry may converge slowly.  This slow convergence can be compensated by use of a preconditioner.  A preconditioner is an approximation to the inverse of the Hessian of the data fidelity term.  For example, consider the WLS data fidelity term.  Then the Hessian (second derivative) is given by $$\Phi''(f|g) = P^*WP.$$  A separable quadratic surrogate (SQS) of this Hessian is given by $$Q = \text{diag}\left( P^*WP\B{1} \right),$$ where $\B{1}$ is a reconstruction where all the values are equal to one.  This can be used as a coarse approximation to the Hessian for use as a preconditioner; since it is a diagonal matrix, its inverse is trivial.  Thus the preconditioned gradient is given by $$Q^{-1}P^*W(Pf-g).$$  The benefits of this preconditioner are shown in Figure \ref{fig:preconditionerEffect}.

\section{Analytic Inversion of the Axial Cone Beam Transform of Symmetric Objects} \label{sec:FBP}

It is well-known that the analytic inversion of the Abel Transform is given by
\begin{eqnarray}
\mathcal{A}^{-1}g(r) &=& -\frac{1}{\pi} \int_r^\infty \frac{g'(y)}{\sqrt{y^2-r^2}} \, dy.
\end{eqnarray}
We wish to determine a similar inversion formula for the cone-beam X-ray Transform for symmetric objects.  To do this, we start with the FBP-type algorithm \cite{FDK} commonly known as FDK.  For a $360^\circ$ acquisition, the FDK algorithm is given by
\begin{eqnarray}
q(\varphi, u, v) &:=& \int_\mathbb{R} h(u-s) \frac{g(\varphi, s, v)}{D^2+s^2+v^2} \, ds \label{eq:rampFilterStep} \\
f(\B{x}) &=& \frac{R}{2\pi}\mathcal{P}^*q(\B{x}), \label{eq:backprojectionStep}
\end{eqnarray}
where $h(s) = -\frac{1}{s^2}$ is the well-known ramp filter.

Since our function, $f$, is invariant for rotations around the axis of symmetry, we may assume that we have projections of the unknown symmetric object around this axis of symmetry.  Then these rotations can be used to re-parameterize the FDK algorithm by a rotation matrix.  This will only change the backprojection step (equation (\ref{eq:backprojectionStep})) of the FDK algorithm.  First we write the X-ray Transform in matrix form
\begin{eqnarray}
\mathcal{P}f(\varphi, u, v) \int f\left( \begin{bmatrix} \cos\varphi & -\sin\varphi & 0 \\ \sin\varphi & \cos\varphi & 0 \\ 0 & 0 & 1 \end{bmatrix} \begin{bmatrix} R-\widehat{l}D \\ -\tau + \widehat{l}u \\ \widehat{l}v \end{bmatrix}  \right) \, d\varphi,
\end{eqnarray}
where we have defined $\widehat{l} := \frac{l}{\sqrt{D^2+ u^2 + v^2}}$ for convenience.

Now let us parameterize our symmetric function by $f(r,z)$, where $r$ is the signed distance from the axis of symmetry and $z$ is along the axis of symmetry.  Then the projections around the tilted axis of symmetry are given by
\begin{eqnarray}
\begin{bmatrix} R-lD \\ -\tau +lu \\ lv \end{bmatrix} = \begin{bmatrix} \cos\alpha & 0 & \sin\alpha \\ 0 & 1 & 0 \\ -\sin\alpha & 0 & \cos\varphi \end{bmatrix} \begin{bmatrix} \cos\varphi & \sin\varphi & 0 \\ -\sin\varphi & \cos\varphi & 0 \\ 0 & 0 & 1 \end{bmatrix} \begin{bmatrix} 0 \\ r \\ z \end{bmatrix} \label{eq:tiltedEquations}
\end{eqnarray}
Thus the backprojection equations for the symmetric cone-beam X-ray Transform are found by solving equation (\ref{eq:tiltedEquations}) for $u$ and $v$ which is given by
\begin{eqnarray}
u(r,z,\varphi) &=& D \frac{r\cos\varphi + \tau}{R-r\sin\varphi\cos\alpha - z\sin\alpha} \label{eq:symmetric_backprojection_u} \\
v(r,z,\varphi) &=& D \frac{-r\sin\varphi\sin\alpha + z\cos\alpha}{R-r\sin\varphi\cos\alpha - z\sin\alpha}. \label{eq:symmetric_backprojection_v}
\end{eqnarray}
Therefore, the FDK reconstruction of a symmetric object with axis of symmetry $(\sin\alpha, 0, \cos\alpha)^T$ may be carried out by calculating equations (\ref{eq:rampFilterStep}, \ref{eq:backprojectionStep}) and replacing equations (\ref{eq:backprojection_u}, \ref{eq:backprojection_v}) by (\ref{eq:symmetric_backprojection_u}, \ref{eq:symmetric_backprojection_v}).  Note that FDK is an quasi-exact reconstruction method (due to the missing cone of frequencies \cite{Leach_NDTE_2021}) and thus so is its adaptation to reconstructing symmetric objects from a single projection.

\section{Implementation of the Forward and Back Projection Algorithms}

When implementing these operators on discrete measurements in software, they can be described as simple matrix-vector multiplication, e.g., $Af = \sum_{j} a_{ij}f_j$ and $A^*g = \sum_i a_{ij}g_i$.  Thus to perform the forward transform and its adjoint, the primary objective is to implement a procedure that calculates the coefficients of the system matrix.

\begin{figure}[h!]
\begin{center}
\includegraphics[width=0.75\textwidth]{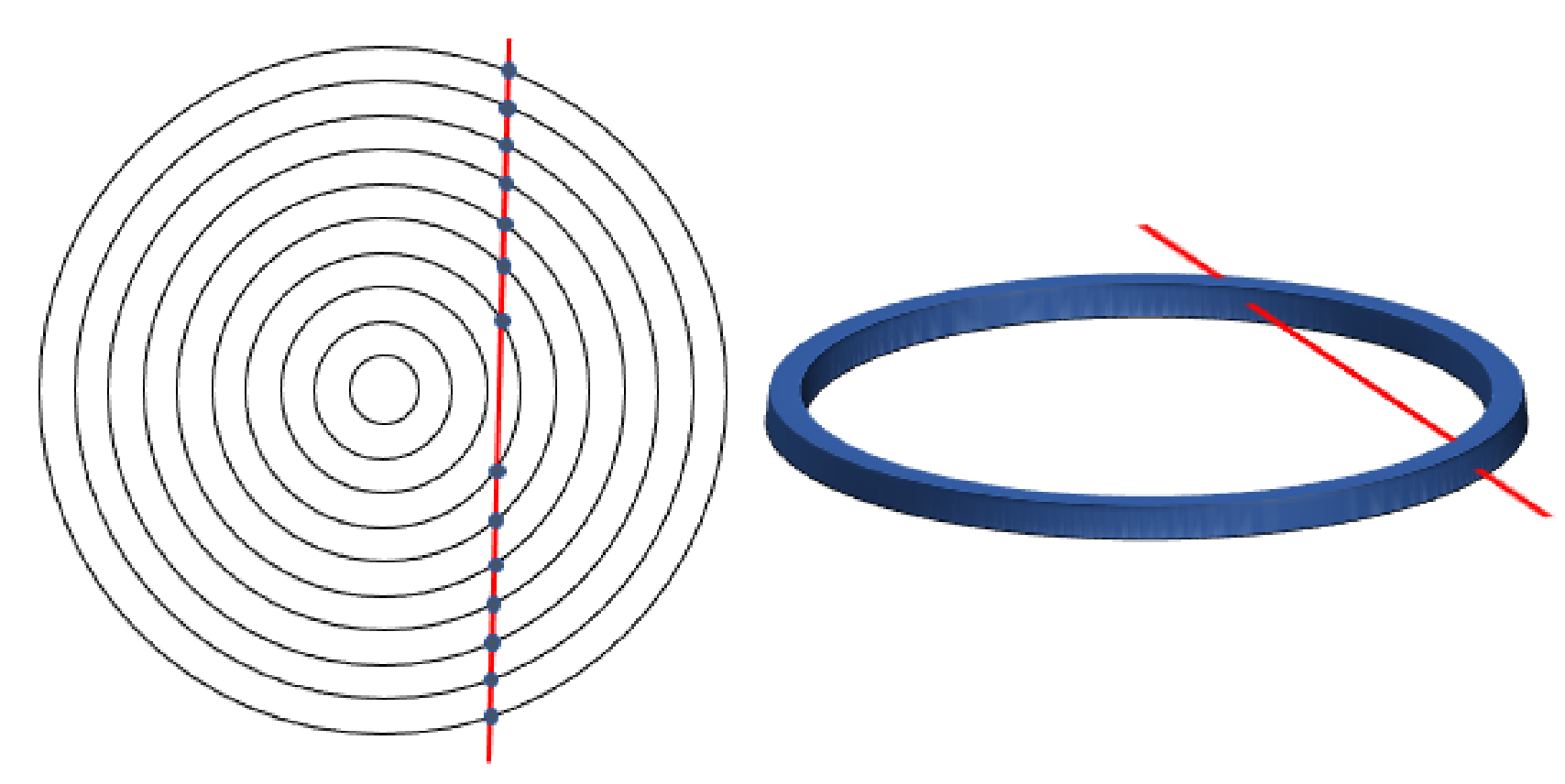}
\vspace{-30 pt}
\end{center}
\caption{Cone-beam projector of symmetric objects is modeled by intersection lengths through annuli.  The image on the left shows the intersection points through the set of annuli perpendicular to the axis of symmetry while the image on the right shows the intersections though a single annuli in three dimensions.} \label{fig:annuliIntersection}
\end{figure}

For the symmetric X-ray Transform transform, the object model is a collection of annuli.  Thus we model the unknown distribution function, $f$, as a collection of annuli of different radii and position along the axis of symmetry, so $a_{ij}$ is the length of intersection between the i-th ray path and the j-th annuli; see Figure \ref{fig:annuliIntersection}.  We assume these annuli are arranged around an axis given by $(\sin\alpha, 0, \cos\alpha)^T$.  We use the cone-beam X-ray Transform parameterization for the measured ray paths at $\varphi = 0$.  The quantity to be computed is then the length of intersection, $l = \sqrt{D^2+u^2+v^2}(t_{max} - t_{min})$, between the line $(R - tD, -\tau + tu, tv )^T$ and the equation for a tilted annulus which is given by
\begin{eqnarray*}
r_{min}^2 \leq \left(x\cos\alpha - z\sin\alpha\right)^2 + y^2 \leq r_{max}^2, \qquad \left| z\cos\alpha + x\sin\alpha - z_0 \right| \leq \frac{h}{2}
\end{eqnarray*}
where $r_{min}$, $r_{max}$ are the inner and outer radii of the annuli, $z_0$ is the position of the annuli along the axis of symmetry, and $h$ is the height of the annuli.  The equations above can be solved using the quadratic formula.

\section{Experimental Validation}

Next, we demonstrate the accuracy and speed of these algorithms.  We shall show the accuracy of our symmetric FBP reconstruction algorithm as described in Section \ref{sec:FBP} and the RWLS algorithm with SQS preconditioner and anisotropic Total Variation regularization \cite{Yu_LineSearch_2006}.

Both CPU and GPU implementations of the forward and adjoint tilted symmetric X-ray Transforms were implemented for parallel-beam and cone-beam geometries in the Livermore Tomography Tools (LTT) \cite{LTTpaper} and LEAP-CT \cite{LEAPpaper} software packages.  These implementations allows transforms to benefit from many model-based iterative reconstruction algorithms, as well as physics-based preprocessing algorithms such as scatter and beam hardening correction.  The GPU-based forward and adjoint calculations of a 2048 $\times$ 2048 cone-beam projection size take about one second on a MacBook Pro with an AMD Radeon Pro 560 GPU.  A list of algorithm computation times are shown in Tables \ref{tab:computationTimes1024} and \ref{tab:computationTimes}.

\begin{table}[h!]
\caption{Computation times for a 1024 $\times$ 1024-pixel projection on a 2017 MacBook Pro Laptop with Radeon Pro 560 GPU} \label{tab:computationTimes1024}
\begin{center}
\begin{tabular}{l|lll}
& Forward Projection & Backprojection & Inversion (FBP) \\
\hline
Parallel-Beam & 0.12 s & 0.12 s & 0.27 s\\
Cone-Beam & 0.16 s & 0.16 s & 0.29 s
\end{tabular}
\end{center}
\end{table}
\begin{table}[h!]
\caption{Computation times for a 2048 $\times$ 2048-pixel projection on a 2017 MacBook Pro Laptop with Radeon Pro 560 GPU} \label{tab:computationTimes}
\begin{center}
\begin{tabular}{l|lll}
& Forward Projection & Backprojection & Inversion (FBP) \\
\hline
Parallel-Beam & 0.55 s & 0.65 s & 1.44 s\\
Cone-Beam & 1.26 s & 0.87 s & 1.57 s
\end{tabular}
\end{center}
\end{table}

\begin{figure}[h!]
\begin{center}
\begin{tabular}{ccc}
Projection & FBP & RWLS \\
\includegraphics[width=0.3\textwidth]{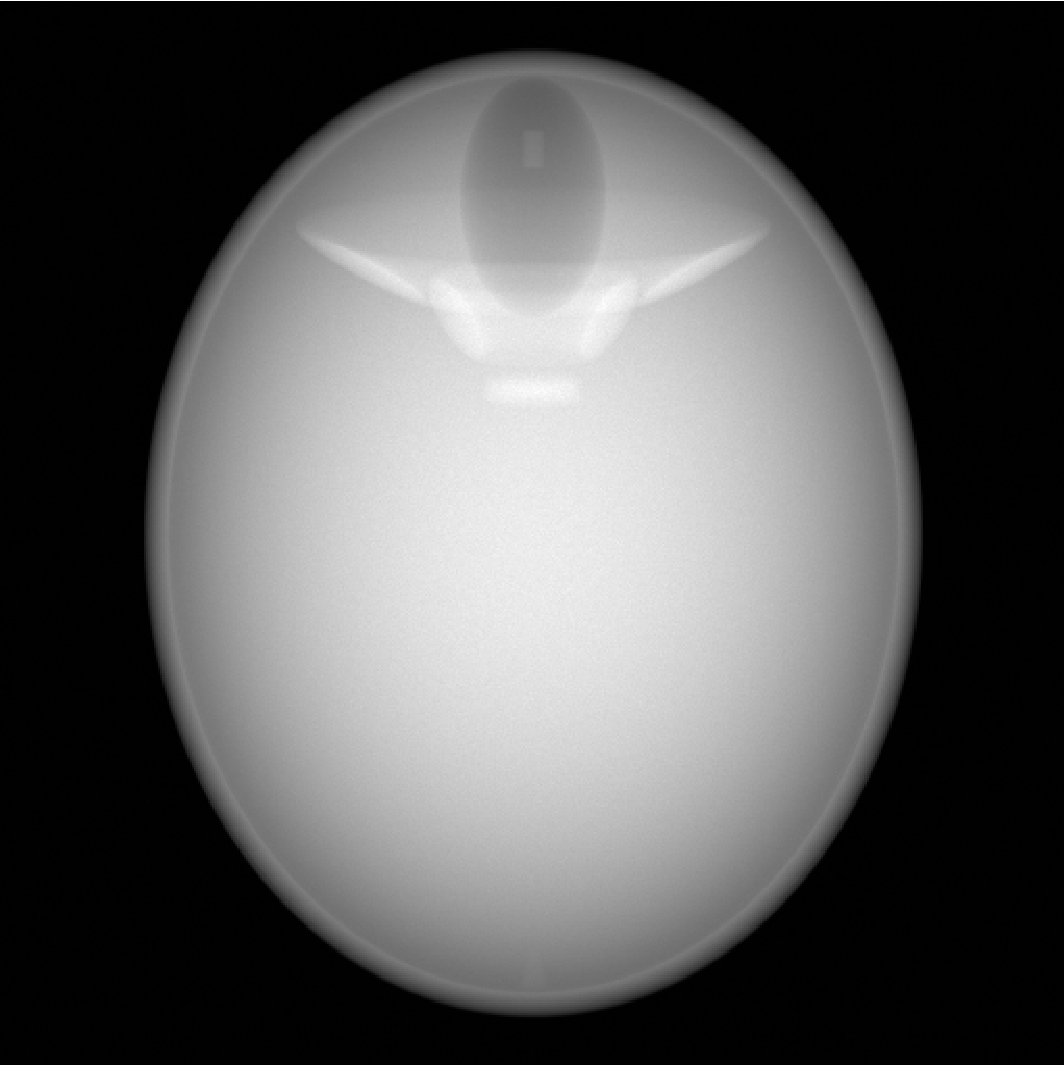}
& \includegraphics[width=0.3\textwidth]{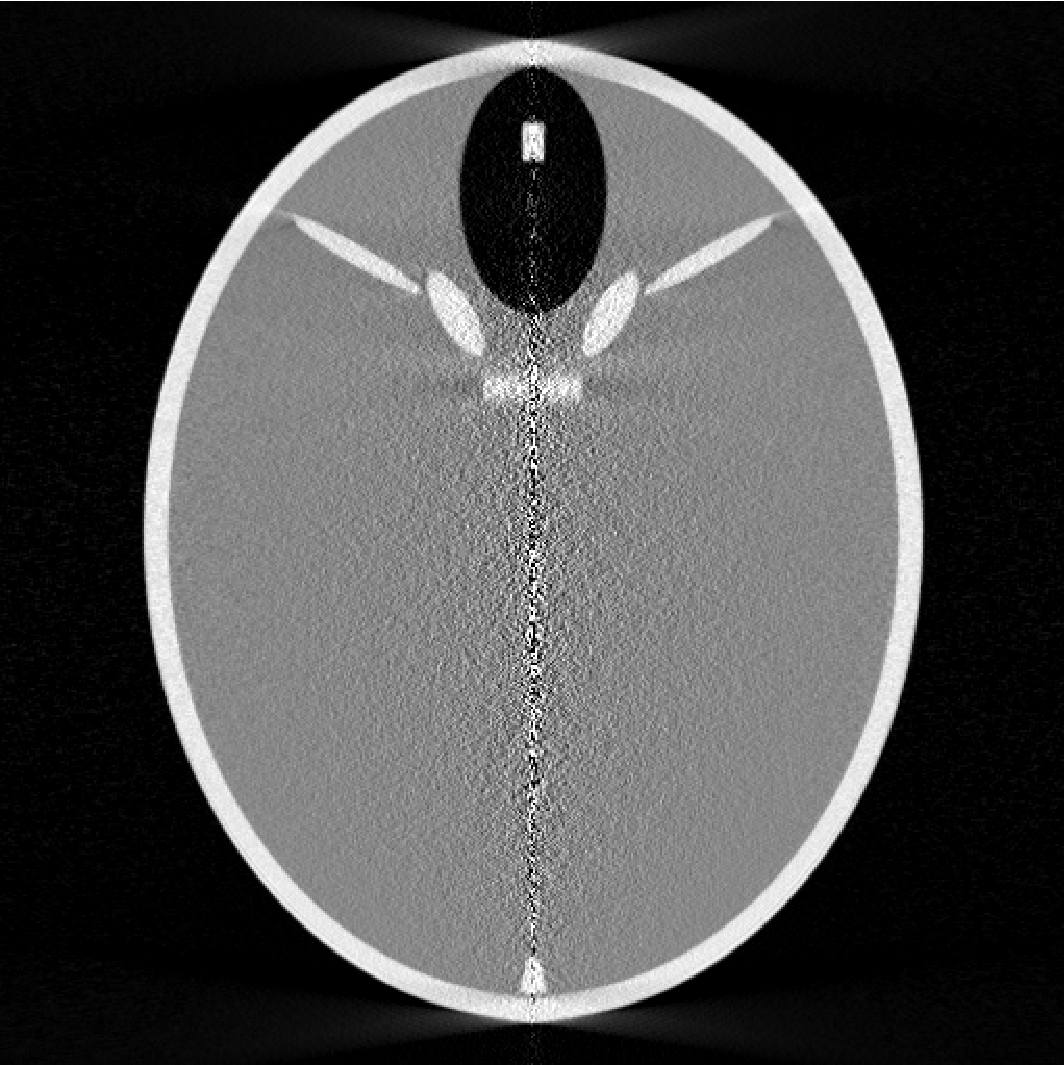}
& \includegraphics[width=0.3\textwidth]{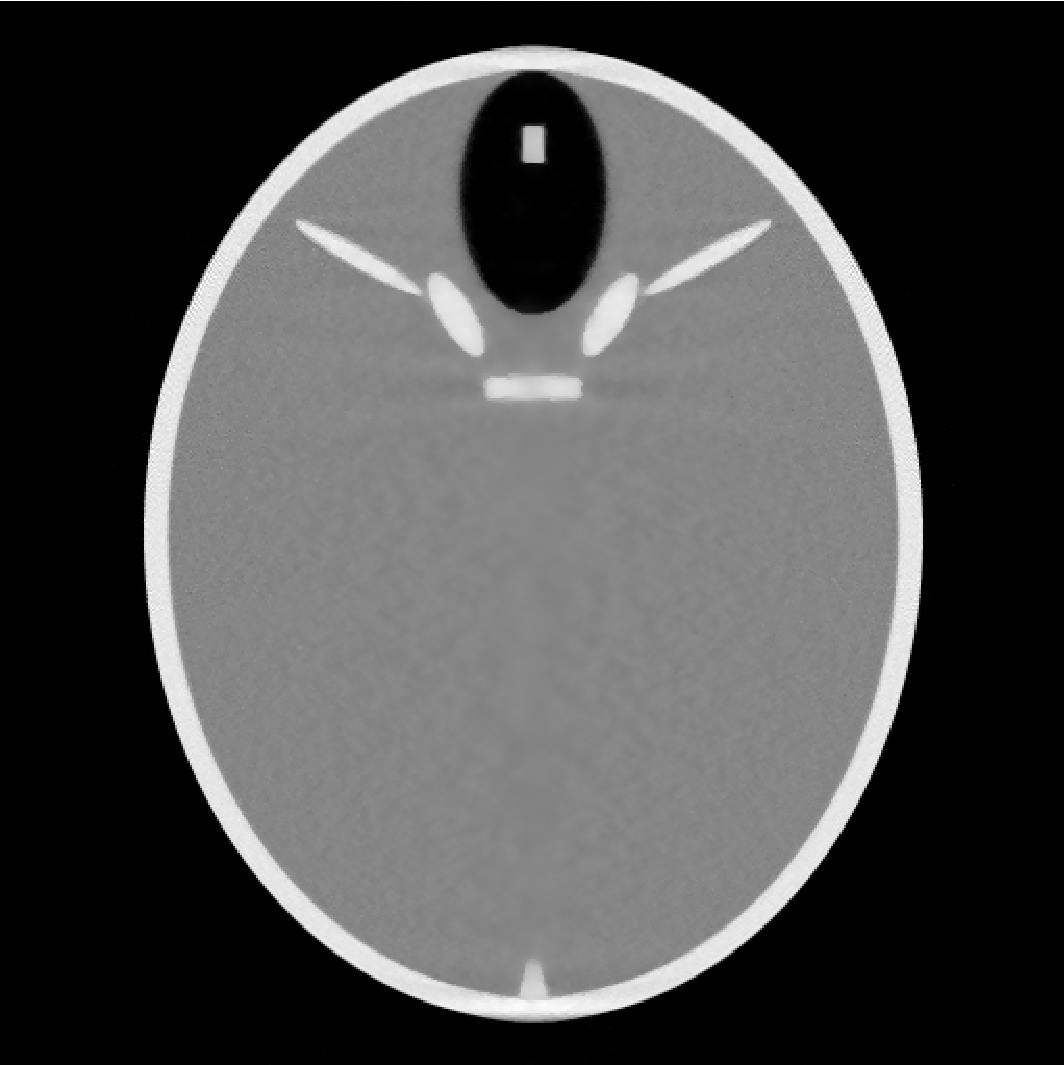}
\end{tabular}
\end{center}
\caption{Reconstruction results for the simulated FORBILD head phantom.} \label{fig:FORBILD}
\end{figure}

We first show results with a simulated dataset that uses a symmetric version of the FORBILD head phantom, a common 3D phantom used to evaluate CT reconstruction algorithms \cite{FORBILD_2012}.  Poisson noise based on the transmission of the measured signal was included in this simulation.  The axis of symmetry parameter, $\alpha$, was ten degrees.  Results are shown in Figure \ref{fig:FORBILD}.  We see that the RWLS reconstruction suppressed cone-beam artifacts and suppressed noise, especially near the axis of symmetry.

\begin{figure}[h!]
\begin{center}
\begin{tabular}{cc}
FBP & RWLS \\
\includegraphics[width=0.3\textwidth]{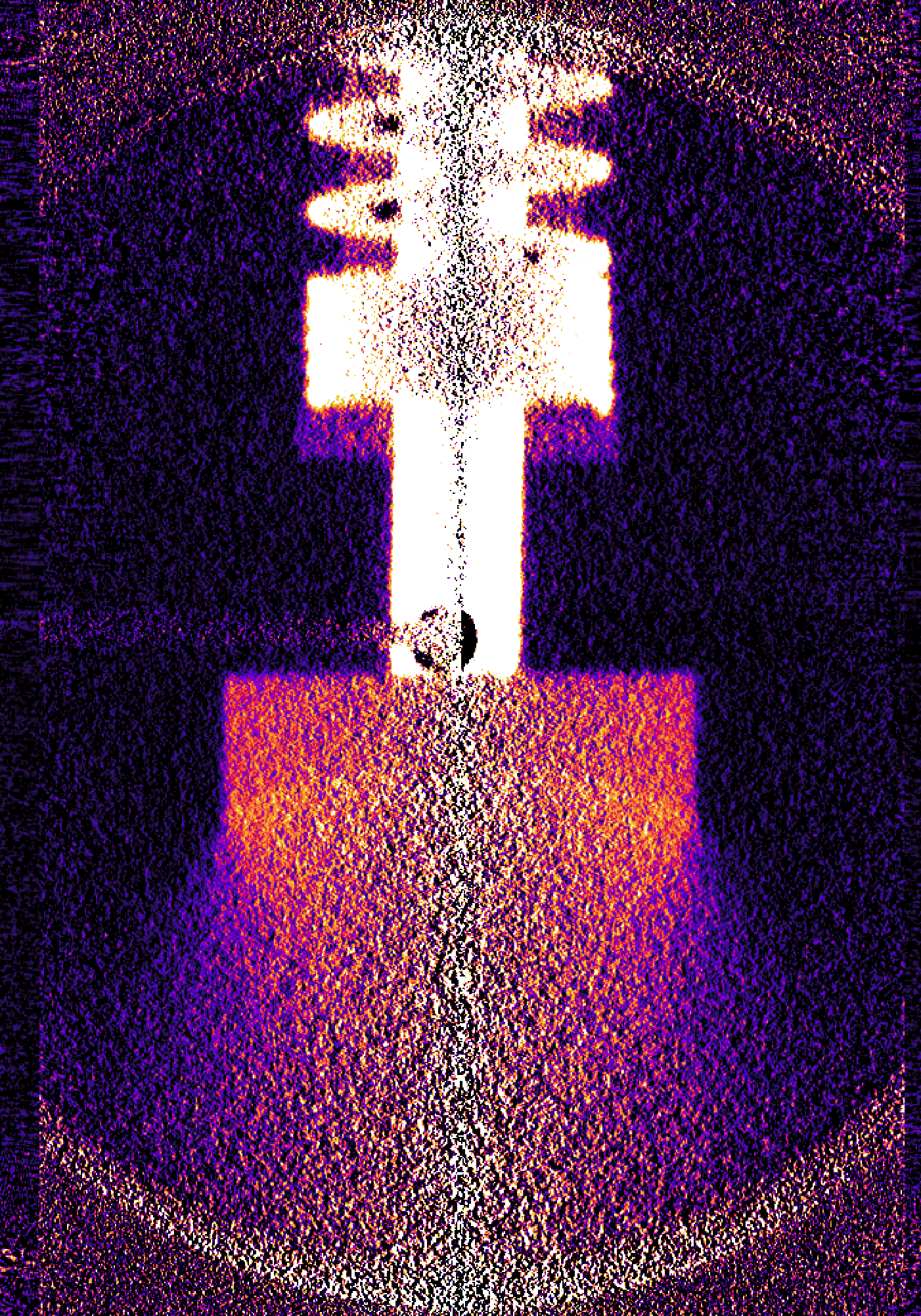}
& \includegraphics[width=0.3\textwidth]{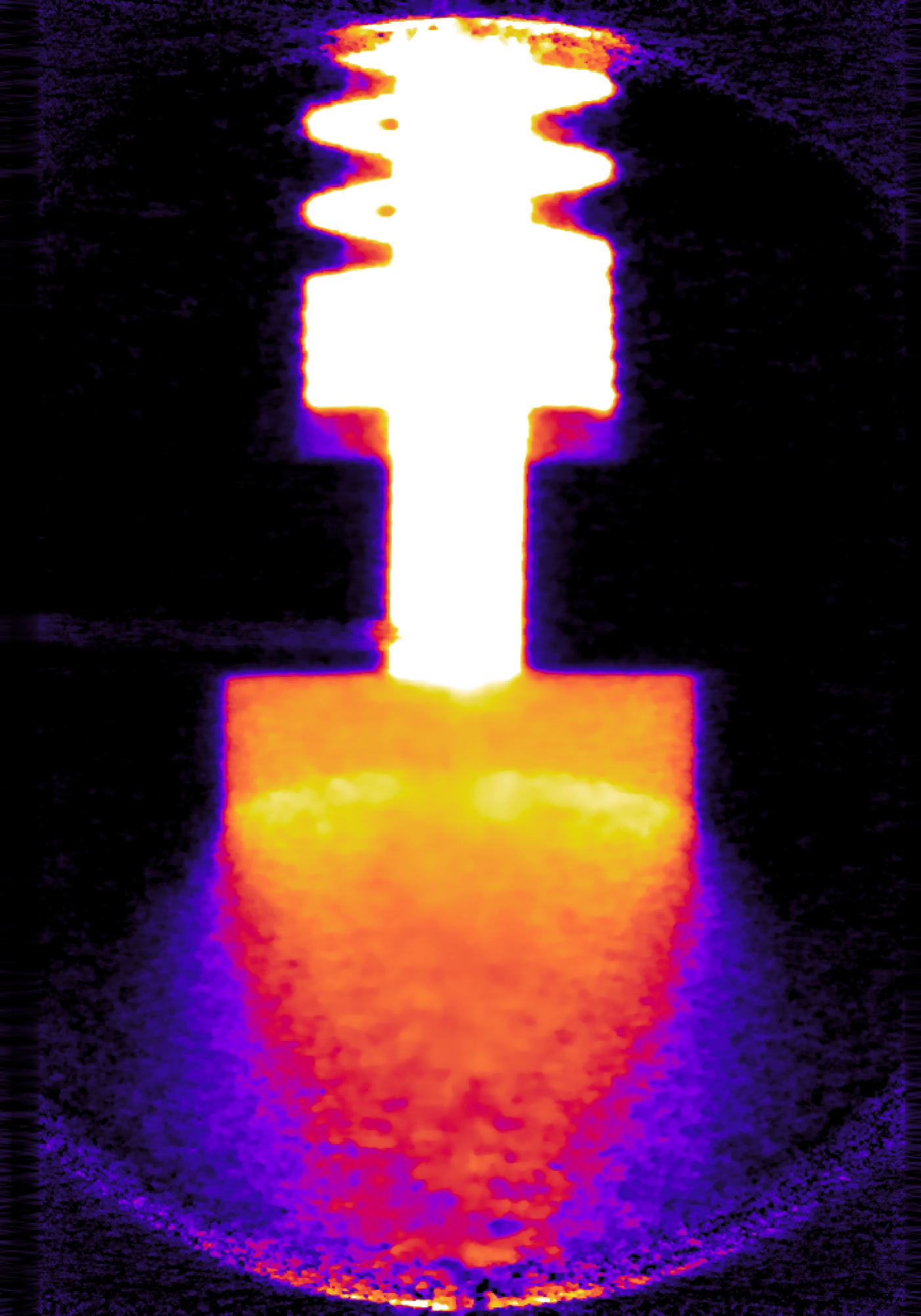}
\end{tabular}
\end{center}
\caption{Reconstruction results the dynamic explosive experiment at ALS.} \label{fig:ALS}
\end{figure}

\begin{figure}[h!]
\begin{center}
\begin{tabular}{ccc}
\includegraphics[width=0.35\textwidth]{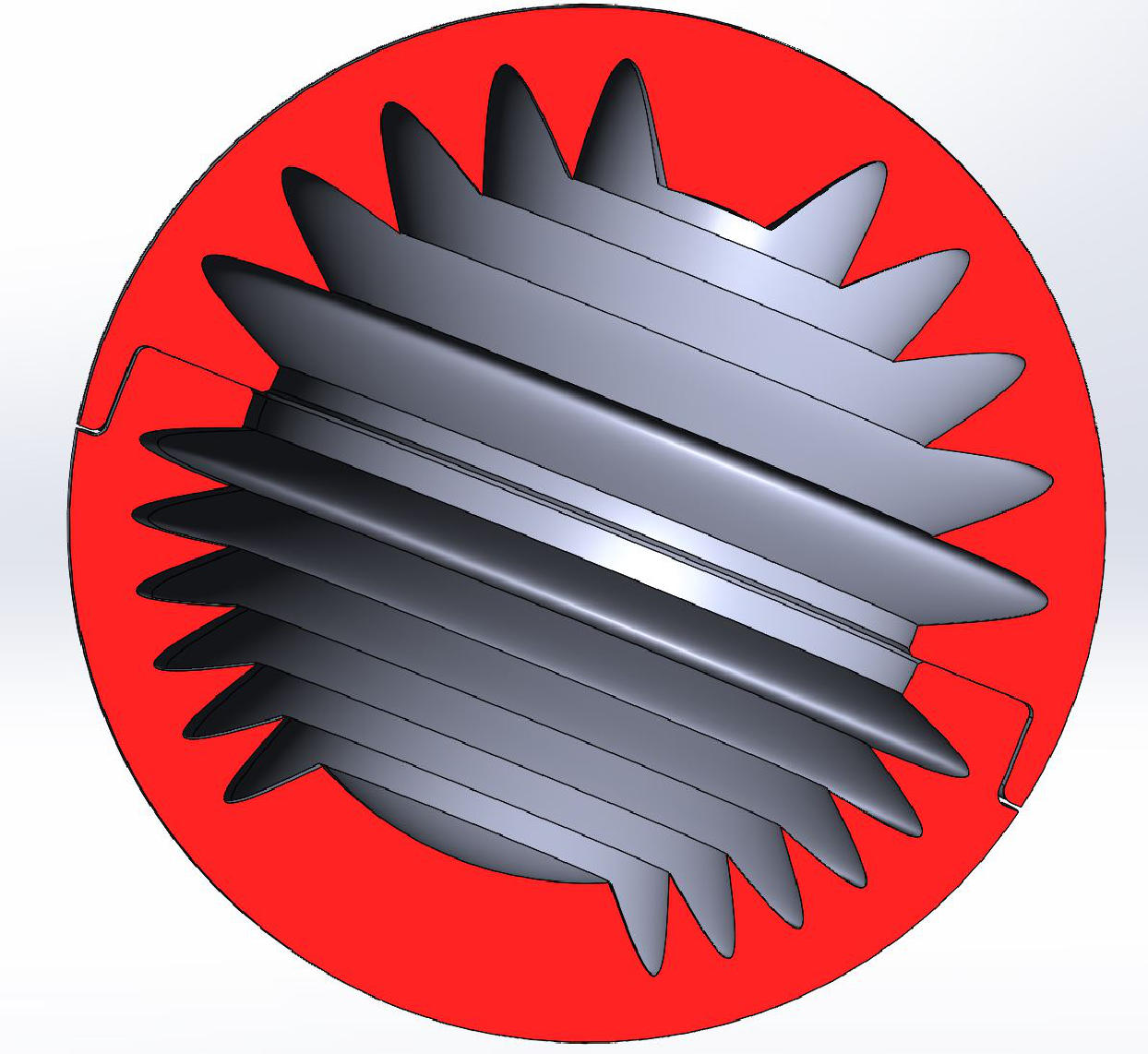}
& \includegraphics[width=0.3\textwidth]{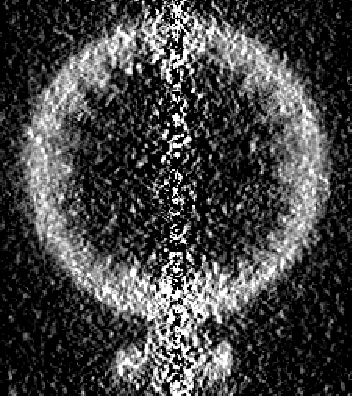}
& \includegraphics[width=0.3\textwidth]{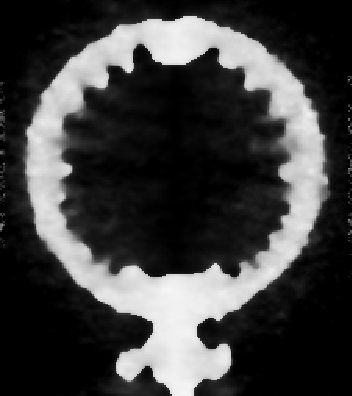}
\end{tabular}
\end{center}
\caption{Reconstruction results using the OSU neutron imaging system, where (left) exemplar drawing, (center) FBP slice, and (right) RWLS slice.} \label{fig:MAS}
\end{figure}

Next we demonstrate our algorithms on experimental data.  These experimental results illustrate the effectiveness of our methods in practical imaging scenarios. The first experimental data was collected of an explosive event captured at ALS \cite{Nielsen2018}.  Results are shown in Figure \ref{fig:ALS}.  The RWLS reconstruction suppressed noise while retaining important image features including the the detonation front.  Noise near the axis of symmetry is especially suppressed.  Since this is a dynamic experiment, it is not possible to reduce noise by taking more frames or integrating longer.  The RWLS reconstruction enables one to make best use of the data.  The second experimental data set was obtained with fast neutrons at the Ohio State University Research Reactor (OSURR) \cite{Oksuz_2020,Oksuz_2021}.  Here the exemplar is a 20 mm diameter metal alloy (95\% Pb, 5\% Sb) sphere with a detailed cylindrically symmetric inner void.  The object was scanned at a tilt angle of five degrees, i.e., $\alpha = 5^\circ$.  The results are shown in Figure \ref{fig:MAS}.  Reconstruction was performed with RWLS, where the regularizer had an anisotropic TV and a histogram sparsity term \cite{fewView_2024}.  The histogram sparsity term is described elsewhere \cite{LTTpaper}.  The noise reduction of the RWLS method is profound.  It is hard to make out the complex inner structures of this object with the FBP reconstruction, but these features are well-resolved in the RWLS reconstruction.

\section{Conclusions and Discussion}

We described the forward, adjoint, and inverse X-ray Transforms operators in both parallel- and cone-beam geometries of objects with an arbitrary axis of symmetry and shown how to implement these in software.  These algorithms are implemented in both the Livermore Tomography Tools (LTT) \cite{LTTpaper} and the LEAP-CT \cite{LEAPpaper} software packages.  This process allows for imaging symmetric objects using realistic system geometries.  For example, much previous work \cite{PyAbel, Asaki_IP_2005, Abramson_SIAM_2008, Howard_SIAM_2016} in reconstructing symmetric objects with a single projection is based on the Abel Transform.  This transform is only appropriate for parallel-beam geometries and objects whose axis of symmetry is perpendicular to the optical axis.  Most imaging systems have a cone-beam geometry, however.  Employing Abel-based methods to cone-beam geometries leads to distortions and reconstruction artifacts.

We showed how to implement Model-Based Iterative Reconstruction (MBIR) algorithms for single view reconstruction of symmetric objects.  Model-based iterative reconstruction can suppress cone-beam artifacts and noise which improves the diagnostic quality of resulting images.  We also showed how to improve the convergence of these methods by employing a Separable Quadratic Surrogate (SQS) preconditioner.

Our methods were validated with both simulated and experimental x-ray and neutron data.  The demonstration of the methods on experimental data is especially compelling because it shows that the algorithms are effective in a real imaging scenarios.  The MBIR methods suppressed cone-beam artifacts and noise especially near the axis of symmetry, while preserving object features.  The MBIR algorithm provided generally superior image quality results to analytic methods.  Although the computational cost is greater for these MBIR algorithms, actual computational times for many problems of interest remain modest.  For example, the fast GPU implementation in LTT enables 2K $\times$ 2K reconstructions in less than a minute on a standard laptop computer.

\section{Acknowledgement}

This work was supported by the US DOE LLNL-LDRD 20-SI-001, and performed under the auspices of the U.S. Department of Energy by Lawrence Livermore National Laboratory under Contract DE-AC52-07NA27344. The document release number is LLNL-JRNL-XXXXXX.  The authors are grateful to Lei Cao and Nerine Cherepy for coordinating the data collection at OSU and Maurice Aufderheide for use of the exemplar in Figure \ref{fig:MAS}.


\end{document}